# Emerging Amines reshape the paradigm of urban atmospheric particle formation

Yongjian Lian[1], Xurong Bai[1], Ruoying Yuan[1], Wenli Xu[1], Yiyang Jiao[1], Hongjun Mao[1], Jianfei Peng[1]*, Shuai Jiang[1]*

[1]Tianjin Key Laboratory of Urban Transport Emission Research & State Environmental Protection Key Laboratory of Urban Ambient Air Particulate Matter Pollution Prevention and Control, College of Environmental Science and Engineering, Nankai University, Tianjin, 300071, China

Corresponding Author: pengjianfei@nankai.edu.cn, shuaijiang@nankai.edu.cn

Table of Contents

**Abstract:**

New particle formation (NPF) contributes to more than half of global aerosol number concentrations, with profound implications for human health and climate change. Observational studies have shown that the frequency of NPF events in urban Beijing during summer exceeds the global average. The prevailing paradigm attributes urban NPF primarily to sulfuric acid–base nucleation involving dimethylamine (DMA). However, recent field measurements in summer urban Beijing have identified several emerging amines emitted from carbon capture processes, including monoethanolamine (MEA), piperazine (PZ), diethanolamine (DEA) and N-methyldiethanolamine (MDEA), in addition to DMA. Here, we systematically evaluate the contributions of sulfuric acid–amine nucleation pathways to urban NPF. We found that emerging amines particularly DEA and PZ, can dominate nucleation pathways under polluted urban conditions, surpassing the contribution of DMA. These findings suggest that the current universal paradigm of urban nucleation should be revisited to explicitly account for the role of emerging amines. Moreover, emerging amine-mediated NPF will become increasingly important in the context of future co-control policies for air pollution and carbon reduction.

**Significance**: Sulfuric acid–dimethylamine (SA-DMA) nucleation is widely recognized as a major driver of new particle formation (NPF) in megacities. However, emerging amines associated with carbon capture technologies have recently been detected in urban atmospheres, and their impacts on atmospheric particle formation remain largely unknown. Here, we provide molecular-level evidence that SA–emerging amines possess substantially stronger nucleation ability than the SA-DMA. Our results show that SA–emerging amines nucleation can dominate summertime urban NPF, contributing up to 80–90% of particle formation in Chinese megacities and helping explain observed NPF events. These findings reveal a previously unrecognized atmospheric consequence of rapidly expanding carbon capture technologies and highlight the need to include emerging amines chemistry in models predicting aerosol formation, air quality, and climate impacts.

**Author contributions**: Y.L. conducted quantum chemical calculations, dynamics simulations and data analysis. X.B., R.Y., W.X. and Y.J. helped write and edit the paper. H.M. improved the terminology expressions of the manuscript. J.P. and S.J. led the conceptualization, methodology development, supervision, and critical revision.

## 1. Introduction

Atmospheric aerosols profoundly affect human health and influence the global climate through aerosol–cloud interactions and aerosol–radiation interactions, including the scattering and absorption of solar radiation(1-5). New particle formation (NPF) contributes more than half of global atmospheric particle number and aerosol mass(6). During NPF, gaseous precursors nucleate to form stable molecular clusters, some of which subsequently grow into cloud condensation nuclei (CCN), thereby affecting the Earth's radiative balance and climate system(7-10). To elucidate the mechanisms governing NPF under diverse atmospheric conditions, extensive field measurements have been conducted in both pristine and polluted environment(11-13). In polluted atmospheres, however, high concentrations of pre-existing particles can suppress NPF by scavenging condensable vapors and newly formed clusters(14-16). Nevertheless, intensive NPF events are still frequently observed in heavily polluted environments(16-20), indicating that the mechanisms driving NPF in polluted atmospheric boundary layers, particularly in megacities, remain incompletely understood.

To better understand nucleation mechanisms in polluted atmospheres, a series of long-term campaigns have been conducted in Chinese megacities using state-of-the-art instrumentation to simultaneously measure nucleating vapors, molecular clusters, and particle number size distributions (PNSDs) (21-25). Previous studies revealed persistently high NPF frequencies during summer in urban Beijing, including ~10.0% in 2004(26), ~25.8% in 2008 (22 NPF days during the entire 85 days)(27) and 16.7% during a 48-day campaign(23). Moreover, a global analysis of 36 observational sites reported summertime NPF frequencies of up to ~45% in urban Beijing, substantially exceeding the global median of 27%(28). Despite the frequent occurrence of NPF events, the underlying triggering mechanisms remain uncertain. Yao et al.(20) and Cai et al.(29, 30) demonstrated that efficient clustering of sulfuric acid ($H_2SO_4$, SA) with atmospheric bases, primarily dimethylamine (DMA), represents the dominant mechanism driving NPF in megacities such as Beijing and Shanghai. However, Yin et al. (31)combining field observations, quantum chemical calculations, and laboratory simulations to demonstrate that ambient DMA concentrations in urban Beijing are insufficient to fully stabilize large sulfuric acid clusters, implying that additional atmospheric bases may play important roles in the nucleation process.

Recently, several emerging amines associated with carbon capture processes were identified during summertime observations in urban Beijing from July to September 2024(25). These compounds include $C_2H_7NO$, $C_4H_{10}N_2$, $C_4H_{11}NO$, $C_4H_{11}NO_2$, and $C_5H_{13}NO_2$, which are likely emitted from carbon capture facilities. Representative species include monoethanol amine (MEA, $C_2H_7NO$), piperazin (PZ, $C_4H_{10}N_2$), 2-amino-2-methyl-1-propan (AMP, $C_4H_{11}NO$), diethanol-amine (DEA, $C_4H_{11}NO_2$), N-methyl diethanol amine (MDEA, $C_5H_{13}NO_2$). Previous studies have shown that these amines can substantially enhance sulfuric acid–base-driven NPF. For instance, Xie et al. combined quantum chemical calculations with the Atmospheric Cluster Dynamics Code (ACDC) model and found that exhibits a stronger cluster formation enhancement potential than methylamine (MA)(32). This unexpected result was attributed to the –OH group in MEA, which can act as both a hydrogen bond donor and acceptor, thereby increasing cluster stability. Ma et al. further showed that PZ can strongly enhance SA-driven nucleation even at parts-per-trillion (ppt) concentrations, with a stronger enhancement effect than DMA, one of the most effective nucleation enhancers identified in previous field and laboratory studies(33). In addition, Wang et al. reported that DEA exhibits a stronger enhancement effect than both DMA and MDEA(34). Despite these advances, the relative contributions of these emerging amines to NPF in typical urban atmosphere remain poorly constrained.

Here, we address this knowledge gap through a systematic investigation of SA-amine systems including MEA, PZ, DEA, MDEA and DMA using quantum chemical calculations combined with the Atmospheric Cluster Dynamics Code (ACDC). We integrate thermochemical analyses of cluster structures with kinetic modeling simulations under typical urban atmospheric conditions. By comparing the nucleation behaviours of SA–emerging amine systems with the conventional SA–DMA system, we quantify cluster stability, formation rates, and their relative contributions to NPF. Our results provide new insights into the molecular mechanisms governing NPF in urban atmospheres and demonstrate that emerging amines associated with carbon capture activities may substantially reshape the prevailing understanding of urban nucleation chemistry.

## 2. Method

### 2.1 Cluster Thermodynamics

To evaluate the stability of cluster formation, we obtained the thermodynamic properties of the SA-amine systems containing MEA, PZ, DEA, MDEA and DMA through quantum chemical calculations. Structures of SA-amine clusters containing MEA, PZ, DMA, DEA and MDEA were taken from previous studies(32-35). To ensure consistency in the level of theory, all structures were reoptimized and their frequencies calculated at the ωB97X-D/6-31++G** level using rigid rotor–harmonic oscillator approximation, with electronic energies further refined at the DLPNO-CCSD(T)/aug-cc-pVTZ level using TightPNO and TightSCF keywords. Geometry optimizations and vibrational frequency calculations were carried out using Gaussian 16(36), while final single-point energy calculations were performed using ORCA 5.0.4(37).

The calculation formula for $G$ is as follows:

$$G = E_{\mathrm{DLPNO}} + G_{\mathrm{thermal}}^{\omega\mathrm{B97X\text{-}D}} \tag{1}$$

$E_{\mathrm{DLPNO}}$ is the single-point energy at the DLPNO-CCSD(T) theoretical level, while $G_{\mathrm{thermal}}^{\omega\mathrm{B97X\text{-}D}}$ is the correction value of Gibbs free energy at the ωB97X-D theoretical level. The free energy of cluster formation ($\Delta G_{\mathrm{ref}}$, kcal/mol) is calculated as the difference between the Gibbs free energy of the cluster and that of the corresponding constituent monomers, using the following formula:

$$\Delta G_{ref} = G_{cluster} - \Sigma G_{monomer} \quad (2)$$

To reflect the thermodynamic energy barrier of cluster formation under real atmospheric concentration conditions, we also calculated the actual Gibbs free energy ($\Delta G_{actual}$) based on standard Gibbs free energy and the measured concentrations of acid and base vapors(38). The expression for $\Delta G_{actual}$ is as follows:

$$\Delta G_{actual}(P_1, P_2, \cdots, P_n) = \Delta G_{ref}\, k_B T \sum_{i=1}^{n} N_i \, In(\frac{P_i}{P_{ref}}) \quad (3)$$

Where, n is the number of components in the cluster, $N_i$ is the number of molecules of type i in the cluster, $P_i$ is the partial pressure of component i in the gas phase, and $P_{ref}$ is 1 atm. $\Delta G_{ref}$ is the Gibbs free energy of formation for the cluster.

### 2.2 Cluster Dynamics

To compare the nucleation abilities of these amines with SA and evaluate their contribution to summer NPF in urban Beijing, we performed cluster dynamics simulations using the Atmospheric Cluster Dynamics Code (ACDC)(38, 39). Evaporation rates were derived from the calculated cluster ΔG values(39), while collision rates were obtained using kinetic gas theory(39). To simulate new particle formation under realistic summer conditions in urban Beijing, we adopted median values of the condensation sink (CS= 0.015 $s^{-1}$) and temperature (T= 306 K) observed during NPF events in summer 2018(23). Monomer concentrations were fixed at the values listed in **Table S1**. Concentrations of SA and emerging amines (MEA, DEA, MDEA, PZ) were based on field measurements reported by Deng et al.(23) and Zhao et al.(25) Due to the lack of direct measurements in summer urban Beijing, DMA concentrations were estimated in this study. Specifically, we used the observed concentration of $C_{2\text{-amines}}$ concentration (~15.38 ppt) proposed by Deng et al.(25) The concentration of $C_{2\text{-amines}}$ is approximated as the concentration of DMA, which has been used in previous NPF studies(23, 29, 31). This approximation is supported by VandenBoer et al., who suggested that DMA is likely the dominant $C_2$-amine in urban atmospheres, often exceeding its isomer ethylamine(40). Based on this, a DMA concentration range of 5–10 ppt was adopted, which likely represents typical summer levels in urban Beijing.

The monomer concentrations listed in **Table S1** were used as inputs for ACDC simulations to estimate the potential formation rates ($J_{potential}$) of SA–amine systems. $J_{potential}$ is defined as the sum of fluxes leading to clusters that are stable enough to grow into larger sizes(41). In ACDC framework, the thermodynamic data for $(SA)_{0–2}(amine)_{0–2}$ clusters were used to calculate fluxes crossing the boundary clusters, including $(SA)_3(amine)_2$, $(SA)_2(amine)_3$ and $(SA)_3(amine)_3$. It should be noted that $J_{potential}$ considers cluster dynamics only up to four molecules and may not fully capture the critical cluster size, potentially leading to an overestimation of formation rates. Therefore, it should not be interpreted as a direct measure of nucleation rate ($J$)(42). However, previous studies suggest that small clusters, such as $(acid)_{1–2}(base)_{1–2}$, represent the rate-limiting steps in nucleation, while larger clusters containing 3–4 acids and bases are generally stable against evaporation(38, 43). Thus, $J_{potential}$ provides a reasonable metric for comparing the ability of different systems to form larger clusters and has been widely used in related studies(41, 44, 45). In addition, we further calculated the nucleation rates ($J_{3×3}$) for $(SA)_x(amine)_y$ (x=0–3, y=0–3) clusters of the top three systems in terms of their contribution to nucleation based on $J_{potential}$. $J_{3×3}$, considers larger clusters, its value will be lower than $J_{potential}$ and closer to the observed nucleation rate in the external field. The boundary conditions used for calculating the $J_{3×3}$ is $(SA)_4(amine)_3$, $(SA)_3(amine)_4$ and $(SA)_4(amine)_4$. These simulations aim to quantify the contribution of emerging amines to new particle formation in urban Beijing under summer conditions.

## 3. Results and Discussion

### 3.1 Cluster Configurations

The five amines investigated here—DMA, MEA, MDEA, DEA, and PZ—differ in the number of hydroxyl and amino substituents, leading to distinct basicities and hydrogen-bonding capabilities toward SA. To evaluate the formation potential of SA–amine clusters, we performed a detailed structural analysis. Proton transfer from SA to the amines was observed in all SA-containing clusters (Fig. S1). The maximum number of proton-transfer events is constrained by the minimum number of two precursor molecules ($n = \min(x, y)$) in $(SA)_x(amine)_y$ ($x=0$–$2$, $y=0$–$2$) clusters. For example, one proton transfer occurs in $(SA)_1$ $(amine)_2$, whereas two proton transfers are observed in $(SA)_2(amine)_2$ clusters (**Table S2**). The amines also exhibit distinct substitution patterns at the nitrogen atom. MDEA contains three alkyl substituents, PZ and DEA contain two, whereas MEA contains one. These structural differences influence amine basicity and proton affinity, thereby affecting cluster formation thermodynamics. Consistent with previous work by Chee et al., which demonstrated that amine basicity correlates strongly with nucleation efficiency(46). The relative basicity inferred from the substitution patterns follows the trend: MDEA > DMA ≈ PZ ≈ DEA > MEA. Increased basicity generally facilitates proton transfer and strengthens acid–base interactions, promoting cluster stabilization.

Hydrogen bonding provides an additional stabilization mechanism that strongly influences cluster stability. Hydroxyl functional groups (-OH) can act as both hydrogen bond donors and acceptors, thereby enhancing cluster stabilization. The investigated amines differ substantially in hydroxyl-group content: DEA and MDEA each contain two –OH groups, MEA contains one, whereas DMA and PZ contain none. These differences lead to pronounced variations in hydrogen-bonding capacity and cluster stabilization. As summarized in **Table S2,** among the $(SA)_1(amine)_1$ clusters, $(SA)_1(DEA)_1$ forms the most hydrogen bonds, followed by $(SA)_1(MDEA)_1$ and $(SA)_1(MEA)_1$, all exceeding that of $(SA)_1(DMA)_1$. For larger clusters, systems containing DEA and MDEA consistently exhibit more extensive hydrogen-bonding networks than the corresponding DMA-containing clusters. A similar trend is observed for pure amine clusters, with hydrogen bond numbers following the order: $(MDEA)_2$ (4) > $(DEA)_2$ (4) > $(MEA)_2$ (2) > $(DMA)_2$ (0) = $(PZ)_2$ (0). These results demonstrate that –OH substituents substantially enhance cluster stabilization through hydrogen-bond formation, potentially altering dominant cluster formation pathways. The thermodynamic consequences of these interactions for cluster formation free energies are discussed in the following section.

### 3.2 Cluster Formation Thermodynamics

Based on the structural analysis above, next evaluated the thermodynamic stability of nucleating clusters in the SA–amine systems. Specifically, the Gibbs free energy of clusters formation ($\Delta G_{ref}$, kcal mol$^{-1}$) were calculated at 278.15 K, 1 atm and the total evaporation rate coefficient ($\Sigma\gamma$, s$^{-1}$) at 306 K as shown in Fig. S2 and S3, respectively. These quantities characterize the thermodynamic and kinetic stability of clusters: more negative $\Delta G_{ref}$ values and smaller $\Sigma\gamma$ indicate that clusters are more likely to form spontaneously through molecular collisions under atmospheric conditions and subsequently contribute to nucleation(47). As illustrated in Fig. S2, the $\Delta G_{ref}$ of SA–amine heterodimers follow the order: $(SA)_1(MDEA)_1$ (-21.79 kcal mol$^{-1}$) > $(SA)_1(DEA)_1$ (-21.19 kcal mol$^{-1}$) > $(SA)_1(PZ)_1$ (-14.15 kcal mol$^{-1}$) > $(SA)_1(DMA)_1$ (-13.52 kcal mol$^{-1}$)> $(SA)_1(MEA)_1$ (-11.76 kcal mol$^{-1}$). This trend is consistent with the differences in amine basicity and hydrogen-bonding capability discussed above. Previous studies by Chee et al and Xie et al. demonstrated that the $\Delta G_{ref}$ of heterodimer correlates with nucleation rates $J_{4\times4}$, allowing its use as a proxy for initial nucleation potential(46-48). Accordingly, SA–MDEA, SA–DEA and SA-PZ systems are expected to exhibit stronger nucleation potential than the conventional SA–DMA system. The stabilizing effect of hydroxyl groups is further reflected in pure amine dimers. The calculated $\Delta G_{ref}$ values for $(DEA)_2$ and $(MDEA)_2$ are -4.06 kcal mol$^{-1}$, -11.65 kcal mol$^{-1}$ respectively, whereas the corresponding values for the other amine dimers are positive. These results further demonstrate that hydroxyl substituents substantially enhance cluster stability through additional hydrogen-bonding interactions.

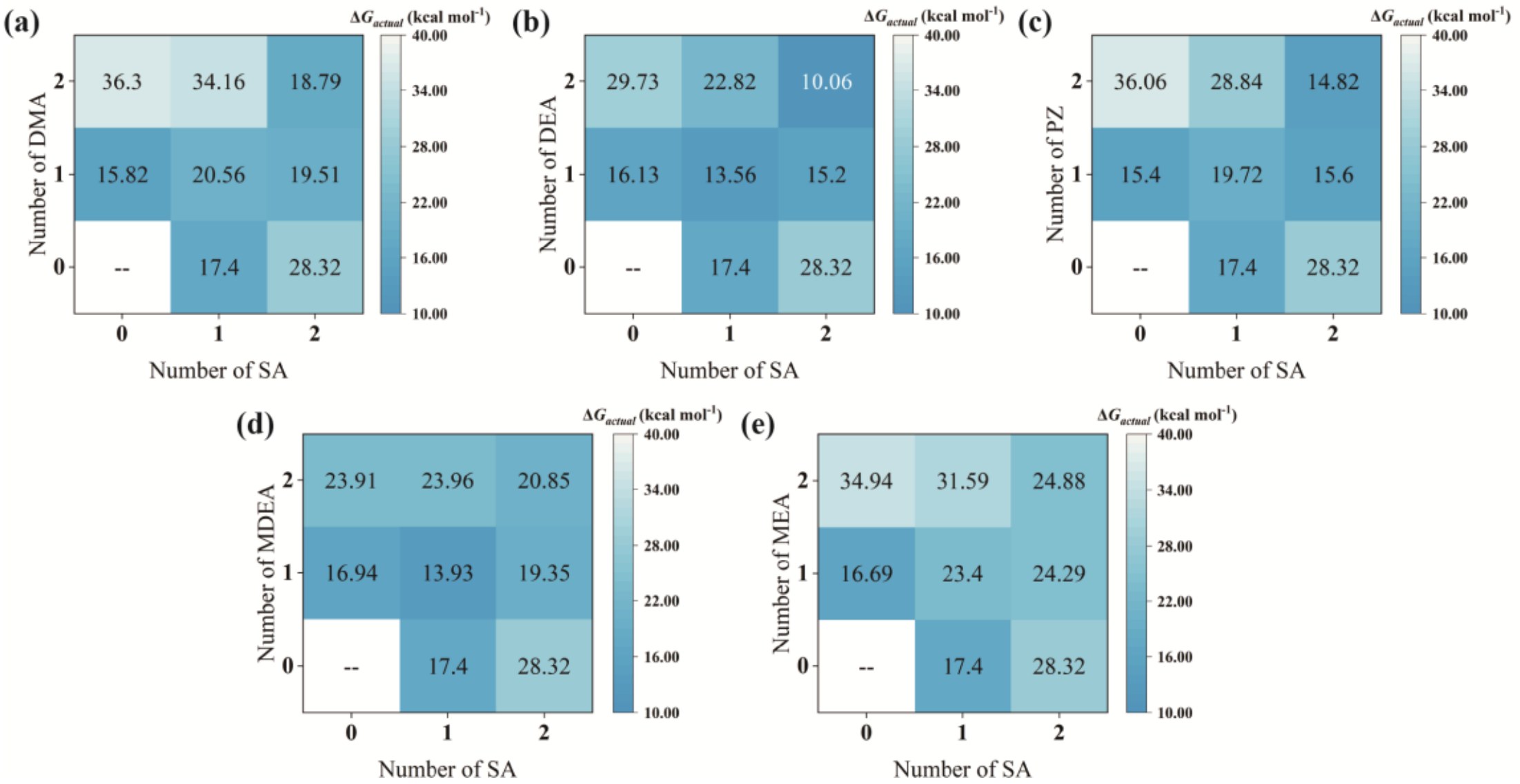

**Figure 1. $\Delta G_{actual}$ of $(SA)_x(amine)_y$ (x=0-2, y=0-2) at 306 K, 1atm, [SA]= $9\times10^6$ cm$^{-3}$ and (a) [DMA]= 7.5 ppt, (b) [DEA]= 3.0 ppt, (c) [PZ]= 8.0 ppt, (d) [MDEA]= 0.8 ppt, (e) [MEA]= 1.2 ppt. The thermodynamics were calculated at the DLPNO-CCSD(T)/aug-cc-pVTZ//ωB97X-D//6-31++G(d,p) level of theory.**

To further evaluate cluster formation under realistic atmospheric conditions, we calculated the actual Gibbs free energies of formation ($\Delta G_{actual}$) for the five SA-amine systems using field-observed precursor concentrations (Fig. 1). Its variation with cluster size provides insight into preferred growth pathways and enables identification of the critical clusters controlling nucleation kinetics. The value of $\Delta G_{actual}$ represents actual formation free energy at the given monomer vapor pressures, the lower value, the easier it is for the corresponding cluster to form under atmospheric concentration conditions. For SA–DMA, SA–DEA, and SA–PZ, the preferred growth pathway follows a diagonal sequence, with $(SA)_1(amine)_1$ identified as the critical cluster. This behaviour indicates stepwise growth may through successive addition of $(SA)_1(amine)_1$ as the mechanism proposed by Cai et al.(30) In contrast, SA–MDEA and SA–MEA systems preferentially follow the pathway of $(SA)_1(amine)_1$ → $(SA)_2(amine)_1$ → $(SA)_2(amine)_2$, with $(SA)_2(amine)_2$ representing the critical cluster. The larger critical cluster size in these systems suggests reduced stability of early-stage clusters, consistent with the thermodynamic trends shown in Fig. S2 and previous findings by Xie et al.(49) The $\Delta G_{actual}$ of critical cluster SA–DEA system exhibits lowest (13.56 kcal $mol^{-1}$), followed by SA–PZ, both of which are lower than that of SA–DMA among all investigated systems. These results indicate that SA–DEA and SA–PZ possess stronger nucleation potential than the conventional SA–DMA system under typical summertime conditions in urban Beijing. The corresponding nucleation rates and atmospheric contributions of these systems are discussed in the following section.

### 3.3 Aerosol Potential Formation Rate

Fig2 (a) illustrates the cluster formation potential ($J_{potential}$) of the SA–amine systems as a function of sulfuric acid (SA) and amine concentrations. To represent summer conditions in urban Beijing, SA concentrations were varied from $1.0\times10^6$ $cm^{-3}$ up to $1.0\times10^7$ $cm^{-3}$, while amine concentrations were constrained to their observed atmospheric ranges at 306 K. Across the investigated SA range, $J_{potential}$ of SA-MEA and MDEA remains below 1 $cm^{-3}$ $s^{-1}$. Even under the highest precursor concentrations, the corresponding formation potentials reached only 0.760 $cm^{-3}$ $s^{-1}$ and 0.839 $cm^{-3}$ $s^{-1}$, respectively. However, SA–DEA, SA–DMA, and SA–PZ systems exhibit substantially higher cluster formation potentials especially SA–DEA. These results indicate that the contributions of SA–DEA and SA–PZ to summertime NPF in urban Beijing are should not to be overlooked. We also analysis the critical clusters controlling nucleation kinetics. As given in Fig2 (b), $\Delta G$ characterizes the energy barrier for a bare sulfuric acid molecule to form a certain cluster. The free energy of a SA is accordingly equal to zero. A positive free energy of $(SA)_1(X)_1$ indicates that the association between $(SA)_1$ and $(X)_1$ needs to overcome an energy barrier ($\Delta G$). SA-DEA system has a free-barrier process, and with an energy barrier of only 2.32 kcal $mol^{-1}$ for SA-PZ, slightly lower than the energy barrier of SA-DMA (3.16 kcal $mol^{-1}$). For SA-MEA and MDEA with $(SA)_2(X)_2$ critical cluster, their energy barriers reach 7.48 kcal mol$^{-1}$ and -11.18 kcal mol$^{-1}$, respectively. Notably, a nearly linear relationship was identified $J_{potential}$ and $\Delta G$ for all $(SA)_1(X)_1$ except SA–MDEA. We have noted the $\Delta G$ of $(SA)_1(MDEA)_1$ is extremely low close to $(SA)_1(DEA)_1$ and far below $(SA)_1(DMA)_1$, but the formation of $(SA)_2(MDEA)_2$ exhibits very unfavorable. From the $\Delta G_{ref}$ in Fig S2 and HBs in **Table S2**, $(SA)_2(MDEA)_2$ has a $\Delta G_{ref}$ that is slightly lower than $(SA)_2(DMA)_2$ but much higher than SA-DEA, and HBs are also lower than SA-DMA and SA-DEA due to the steric hindrance effect of methyl substituents on N atoms, means lower cluster stability. In addition, it is necessary to consider the lower atmospheric concentration of MDEA (0.3~1.16 ppt) than other amines, which makes cluster formation more unfavorable. These results indicates that heterodimer stability alone is insufficient to fully predict nucleation potential, and that the thermodynamic properties of larger clusters must also be considered to comprehensively evaluate atmospheric nucleation efficiency.

In addition, we found $J_{potential}$ of SA–PZ system ranges from 4.070 to 410.497 cm$^{-3}$ s$^{-1}$ at low PZ concentration (5.21 ppt) and from 9.807 to 981.724 cm$^{-3}$ s$^{-1}$ at high PZ concentration across the same SA range. Under comparable conditions, SA–DMA yields lower values of 1.517~152.967 cm$^{-3}$ s$^{-1}$ at 5.0 ppt DMA and 3.528~353.441 cm$^{-3}$ s$^{-1}$ at 10.0 ppt DMA. Most strikingly, the SA–DEA system exhibits formation potentials that substantially exceed those of both SA–DMA and SA–PZ. At a representative DEA concentration of 3.0 ppt, which is lower than the maximum observed concentrations of DMA and PZ, SA–DEA still produces $J_{potential}$ values of 102.541~9715.92 cm$^{-3}$ s$^{-1}$, exceeding those of both SA–PZ and SA–DMA by more than an order of magnitude. Even at extremely low DEA concentrations (0.16 ppt), the corresponding $J_{potential}$ remains high (0.765~166.749 cm$^{-3}$ s$^{-1}$), approximately 1000 times larger than those of the SA–DMA system under comparable precursor concentrations (0. 001~0.229 cm$^{-3}$ s$^{-1}$). These results demonstrate that SA–DEA nucleation can surpass the traditionally dominant SA–DMA pathway under summertime conditions in urban Beijing. It should be noted that $J_{potential}$ are generally higher than experimentally observed nucleation rates because larger clusters and subsequent growth losses are not explicitly included in the simulations. Nevertheless, $J_{potential}$ provides a robust metric for comparing the relative nucleation efficiencies of different precursor systems. Herein, the three systems with the highest cluster formation potentials such as SA–DEA, SA-PZ and SA-DMA were selected for further evaluation of their contributions to atmospheric nucleation rates.

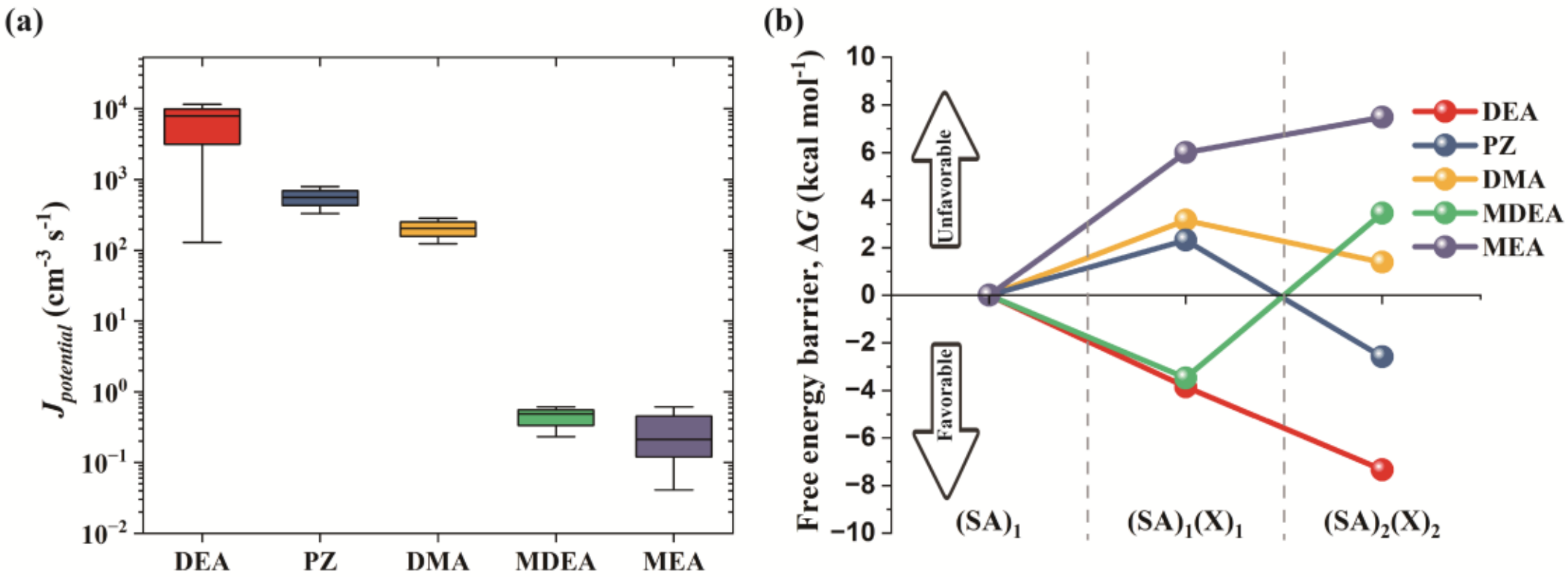

**Figure 2. (a) Aerosol potential formation rates ($J_{potential}$, cm$^{-3}$ s$^{-1}$) at T = 306 K, 1atm, CS = 0.015 s$^{-1}$, [SA] =9 × 10$^{6}$ cm$^{-3}$, [MEA] =0.6-1.94 ppt, [PZ] =5.21-10.9 ppt, [DEA] =0.16-5.82 ppt, [MDEA]=0.38-1.16 ppt, [DMA]=5.0-10.0 ppt. (b) Free energy barrier ($\Delta G$, kcal mol$^{-1}$) for $(SA)_1$, $(SA)_1(X)_1$, $(SA)_2(X)_2$ at T = 306 K. The typical concentrations of gaseous precursors in summer urban Beijing were selected as [SA] =9 × 10$^{6}$ cm$^{-3}$, [MEA]=1.2 ppt, [PZ]=8.0 ppt, [DEA]=3.0 ppt, [MDEA]=0.8 ppt, [DMA]=7.5 ppt.**

To connect the simulation results with real atmospheric conditions, we compared the simulated nucleation rates ($J_{3\times3}$) of the SA-DEA, SA-PZ, and SA-DMA systems with field-observed NPF rates in summertime urban Beijing as given in Fig 3(a). The SA–DMA mechanism alone is insufficient to reproduce the observed nucleation rates, even under the highest SA and DMA concentrations. In contrast, incorporating the SA–DEA and SA–PZ mechanisms substantially improve agreement with field observations and reasonably reproduces the observed nucleation intensities in urban Beijing during summer. Notably, under relatively low SA concentrations (3×10$^{6}$ cm$^{-3}$), the SA–DEA system at 3.0 ppt DEA produces higher nucleation rates than the SA–PZ system at 10.9 ppt PZ. These findings indicate that the SA–DEA pathway may represent a dominant nucleation channel during summertime NPF events in Beijing.

To further elucidate the origin of these differences, we investigated the dynamic formation mechanisms of SA-amines systems at 306 K under a condensation sink (CS) 0f 0.015 $s^{-1}$ and representative summertime precursor concentrations in urban Beijing (Fig 3b). The initial growth step is same for SA-DEA, SA-PZ and SA–DMA systems, involving formation of the $(SA)_1(amine)_1$ heterodimers. However, the subsequent growth pathways differ substantially among the systems. For SA-DMA, cluster growth primarily proceeds through sequential addition of one SA molecule, followed by one DMA molecule, and finally by collision with another $(SA)_1(DMA)_1$ cluster. This mechanism differs from that proposed by Cai et al. for wintertime Beijing, where growth predominantly occurs through successive addition of $(SA)_1(DMA)_1$ cluster ($A_nD_n$+$A_1D_1$)(30). Our simulations further show that, as temperature decreases from 306 to 266 K (Fig. S6), the contribution of the pathway involving collision between two $(SA)_1(DMA)_1$ cluster to form $(SA)_2(DMA)_2$ increased to ~45%. This temperature dependence indicates that summertime SA–DMA nucleation is strongly constrained by the limited stability of the of $(SA)_1(DMA)_1$ heterodimer. In contrast, owing to the exceptional stability of the $(SA)_1(DEA)_1$ and $(SA)_2(DEA)_2$ clusters, dominant SA–DEA growth mechanism involves only two major growth steps: adds $(SA)_1(DEA)_1$and $(SA)_2(DEA)_2$. This highly efficient growth pathway likely explains the remarkably strong nucleation capability of the SA–DEA system. The enhanced stability of these DEA-containing clusters is attributed to the presence of two hydroxyl groups in DEA, which strengthen hydrogen-bonding interactions and suppress evaporation. Consistently, evaporation rate analysis supports this interpretation: lower $\Sigma\gamma$ values indicate greater cluster stability and reduced evaporation. As shown in Fig. S3, key clusters such as $(SA)_1(DEA)_1$ and $(SA)_2(DEA)_2$ maintain exceptionally low evaporation rates ($\Sigma\gamma < 10^{-3}\ s^{-1}$) even at the elevated summertime temperature of 306 K. For SA-PZ, the dominant growth pathway proceeds through stepwise addition of either SA or PZ molecules, consistent with the mechanism reported by Ma et al(33). Unlike the SA-DMA system, which is strongly limited by the stability of the $(SA)_1(DMA)_1$ cluster, the SA-DEA and SA-PZ systems are primarily constrained by the atmospheric availability of SA and base molecules. Consequently, under summertime conditions, sufficiently high concentrations of SA and DEA or PZ can strongly enhance NPF in polluted urban environments.

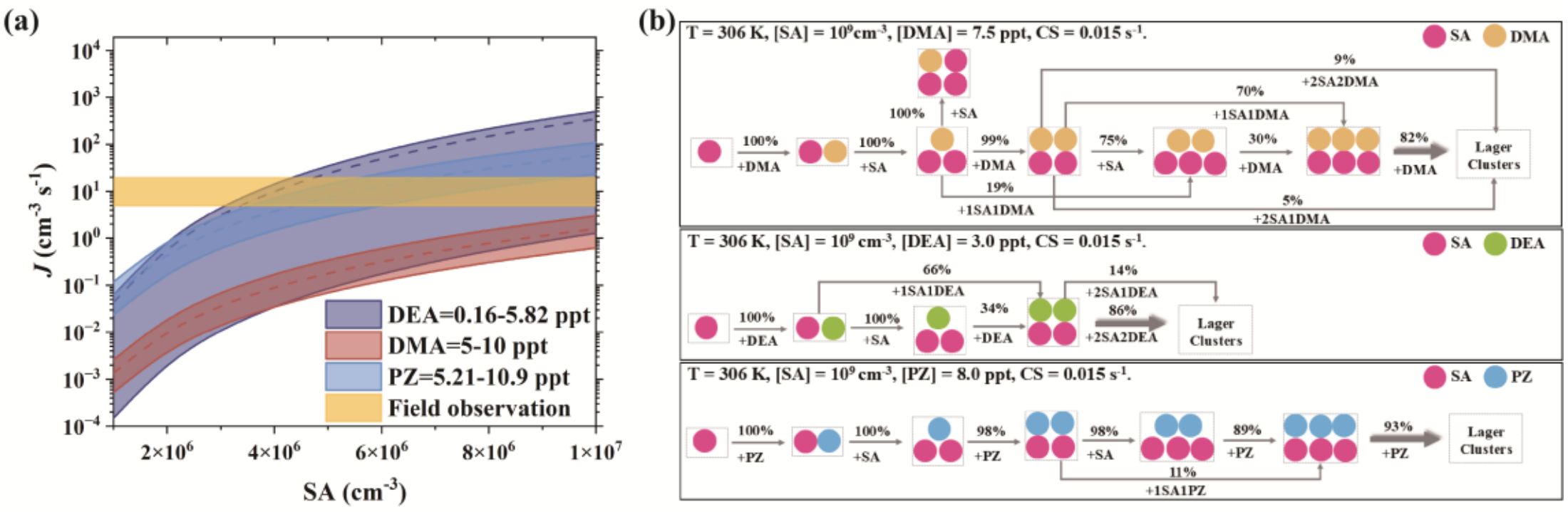

**Figure 3. (a) Aerosol formation rates ($J$, cm$^{-3}$ s$^{-1}$) at T = 306 K, 1atm, CS = 0.015 s$^{-1}$, [SA] =$10^6$-$10^7$cm$^{-3}$, [PZ] =5.21-10.9 ppt, [DEA] =0.16-5.82 ppt, [DMA] =5.0-10.0 ppt. The dashed lines represent the concentration of amine is median value: [DMA]= 7.5 ppt, [DEA]= 3.0 ppt, [PZ]= 8.0 ppt. (b) Cluster formation pathways for the SA−DMA, SA−DEA, SA−PZ system. To simplify the growth path, paths with contributions below 5% were artificially removed.**

## 4. Atmospheric Implications

To comprehensively evaluate the contribution of emerge amines to NPF events in urban areas, we selected a sufficiently broad range of precursor concentrations, covering typical atmospheric conditions in summer and winter in polluted urban areas of China ([SA]= $1\times10^5$-$1\times10^8$ cm$^{-3}$and [amine]=0.01-100 ppt) for the input of ACDC. Detailed simulation parameters are provided in **Table S4**. As illustrated in Fig. 4, we found that in summer, the contribution of emerge amines to nucleation rate is absolutely dominant (90~100%), but in winter, the importance of emerge amines slightly decreases but still maintains its importance, except when the concentration of DMA is extremely high ([amine]=100 ppt), where the contribution of SA-DMA (~60%) begins to surpass that of emerge amines. In addition, we also evaluated the importance of emerge amines under atmospheric conditions in several typical urban areas, such as Beijing, Shanghai, and Wangdu in Hebei. The observation data comes from previous work(20, 23, 50) as given in **Table S5**. It can be observed that during the summer and winter seasons in Beijing and Shanghai, the contribution of emerge amines to nucleation rate remains dominant at 80-90%. However, in the urban area of Wangdu, the contribution of emerge amines slightly decreases but is still comparable to SA-DMA. These results suggest that emerge amines may play a dominant role in summer NPF in urban, potentially surpassing DMA. Given the increasing use of carbon capture technologies in urban areas, emerge amines emissions may be widespread throughout the year in major cities. However, this distribution remains uncertain. Further measurements and studies, SA-base driven clusters, are needed to better constrain the contribution of emerge amines to atmospheric nucleation.

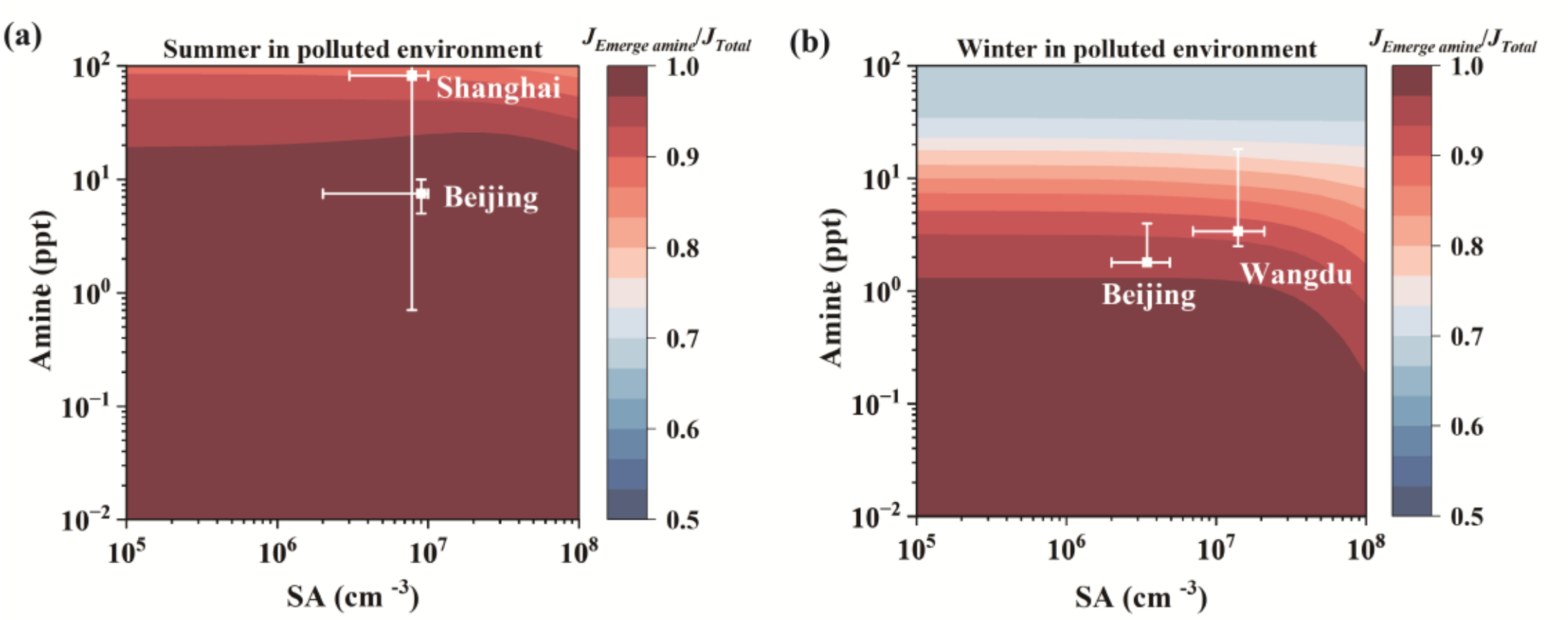

**Figure 4. The illustrative figures for the contribution ratio of new amine to nucleation rate ($J_{Emerge\ amine}/J_{Total}$, $J_{Emerge\ amine}= J_{SA\text{-}DEA}+ J_{SA\text{-}PZ}$, $J_{Total}= J_{SA\text{-}DEA}+J_{SA\text{-}PZ} +J_{SA\text{-}DMA}$) in various conditions, where $J_{SA\text{-}DEA}$, $J_{SA\text{-}PZ}$ and $J_{SA\text{-}DMA}$ represent the simulated nucleation rates (J3×3) of the SA-DEA, SA-PZ, and SA-DMA, respectively. (a) Summer (T= 300 K) and (b) Winter (T= 278 K) in polluted environment (CS= 2.7 × $10^{-2}$ $s^{-1}$). T and CS are the temperature and the condensation sink coefficient, respectively. The open markers and error bars indicate the median values and approximate ranges, respectively, of the effective sulfuric acid and amine concentrations in Beijing, Shanghai and Wangdu. The Beijing data were reported by Deng et al. (2020)(23). The Shanghai data were reported by Yao et al. (2018)(20). The Wangdu data were reported by Jin et al. (2025)(50). Note that for the concentrations of PZ and DEA, we assume that its concentration is consistent with the observation range of DMA, so that we can**

**fairly compare the contributions of these amines to the nucleation rate. As a result, there are potential uncertainties in the results for PZ and DEA.**

Carbon-capture-related amine emissions are expected to evolve with the future deployment of amine-based CCUS technologies, solvent composition, and emission-control efficiency. To evaluate the potential influence of future CCUS expansion on urban nucleation chemistry, we introduced a dimensionless CCUS influence index ($I_{CCUS}$), defined as the ratio between future ambient concentrations of emerging amines and the median concentrations observed in urban Beijing during 2024. Technology-deployment scenarios were represented by ($I_{CCUS}$= 0.3, 1, 3, and 10) corresponding to controlled-CCUS, present-day, near-future, and high-CCUS-influence conditions, respectively. In addition, Zhu et al. proposed that gasoline-powered and dieselpowered vehicles contribute 82.5 % to $C_{2\text{-amines}}$ (primary DMA) emissions in urban Beijing. Herein, to simultaneously consider the impact of future pollution control and emission reduction policies, we have also taken corresponding measures for the concentration of DMA as given in Table S6. These scenarios are not intended as explicit emission forecasts, but rather as sensitivity-based regime analyses designed to evaluate how increasing anthropogenic amine influence may shift urban sulfuric-acid-driven nucleation from the conventional DMA-dominated regime toward emerging-amine-dominated pathways.

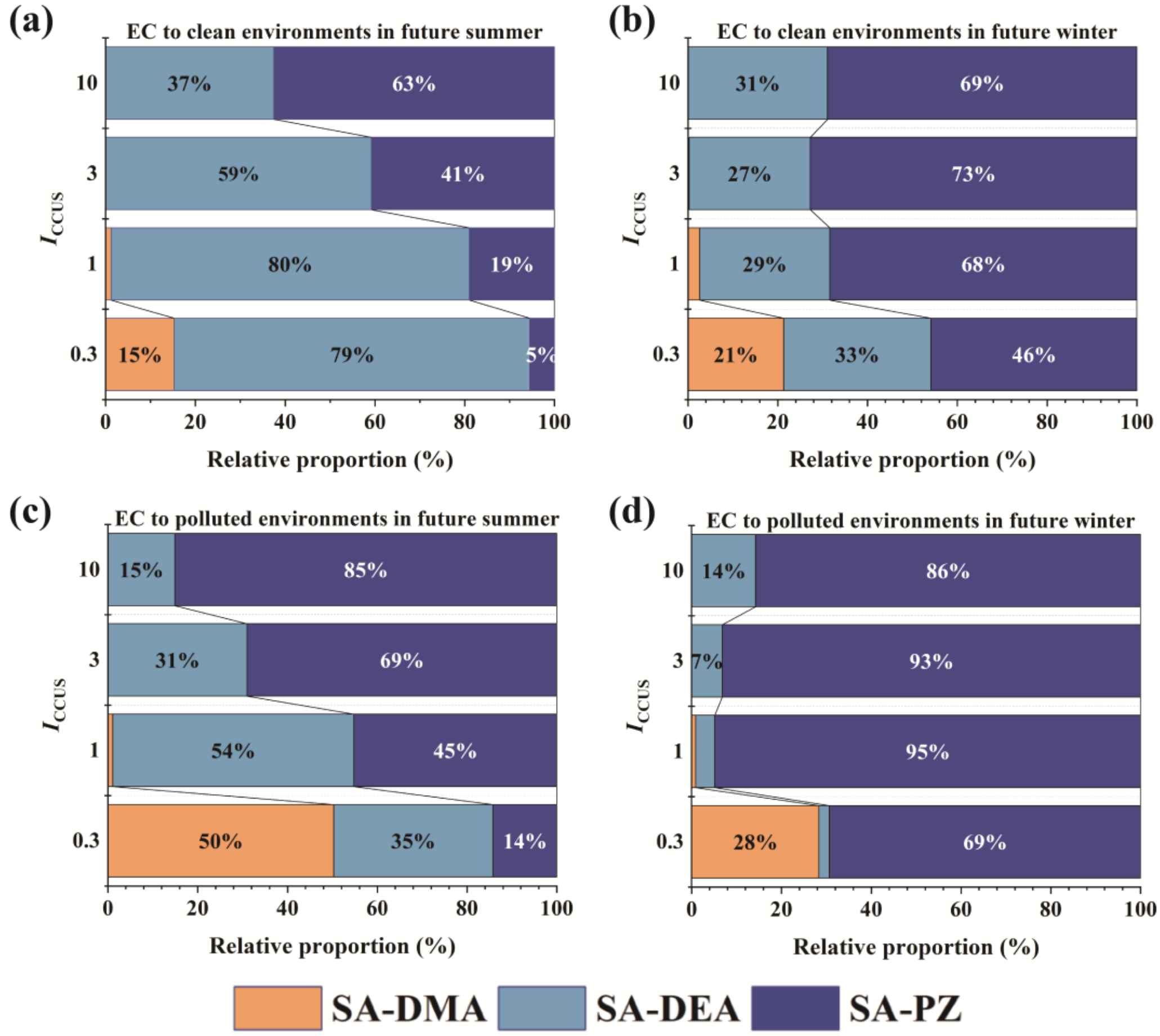

**Figure 5. Estimated contribution (EC, %) of SA–DMA, SA–DEA, and SA–PZ nucleation under clean and polluted environmental conditions in future (a) summer and (b) winter scenarios across different levels of CCUS technology deployment and air pollution control. Note that paths with contributions below 5% were artificially removed. All simulations above assume constant SA concentrations across scenarios.**

As illustrated in Fig. 5, under the combined influence of increasing deployment of CCUS technologies and progressively strengthened air pollution control from present to future scenarios, atmospheric DMA concentrations continuously decrease, whereas concentrations of emerging amines gradually increase. Under extensive CCUS deployment ($I_{\text{CCUS}}$ =10), DMA emissions become negligible compared with those of emerging amines, leading nucleation involving SA and emerging amines to dominate essentially all SA–amine-driven nucleation pathways. In addition, under both clean and polluted summer conditions, the dominance of DEA in SA–emerging amine nucleation gradually weakens with increasing CCUS deployment and is eventually surpassed by PZ. Although the thermodynamic analysis above indicates that SA–DEA nucleation is more favorable than SA–PZ nucleation, but when $I_{CCUS}$ =10, the atmospheric concentration of PZ becomes extremely high (~1.97 × $10^9$ $cm^{-3}$). Under such high-temperature and high-PZ-concentration conditions, despite the thermodynamic advantage of DEA, the limited SA concentration (~9.0 × $10^6$ $cm^{-3}$) substantially increases the collision probability between SA and PZ molecules, thereby driving SA to nucleate preferentially with PZ. This phenomenon becomes even more pronounced under clean and polluted winter conditions, where SA–PZ nucleation persistently dominates the nucleation process. Nevertheless, the contribution of DEA gradually increases with increasing $I_{CCUS}$, likely because lower temperatures suppress cluster evaporation and partially weaken the influence of concentration differences on cluster formation. Moreover, under identical environmental conditions, the contribution of PZ-mediated nucleation is consistently higher in winter than in summer, whereas DEA-mediated nucleation exhibits the opposite seasonal trend, with greater contributions in summer. Because, SA-DEA system has more favorable thermodynamic properties than SA-PZ.

When CCUS deployment decreases and air pollution control becomes weaker, the contribution of DMA-mediated nucleation increases markedly, particularly under polluted summer conditions (EC = 50%). These results suggest that, under future scenarios characterized by widespread CCUS implementation and enhanced air pollution control, the dominant role of DMA in urban nucleation will likely be replaced by emerging amines, leading to complex nucleation mechanisms. In the future, as emissions of conventional anthropogenic pollutants from agriculture and industry continue to decline while emissions of emerging amines increase because of new technologies, emerging amine-mediated NPF is expected to exert an increasingly important influence on global climate forcing.

## ASSOCIATE DCONTENT

The Supporting Information is available free of charge

Identified global minimum configurations of $(SA)_x(amines)_y$ (x=0-2, y=0-2) at the DLPNO-CCSD(T)/aug-cc-pVTZ//ωB97X-D/6-31++G(d,p) level of theory; $\Delta G$ of $(SA)_x(amine)_y$ (x=0-2, y=0-2) at 278 K, 1atm; The total evaporation rate coefficient $\sum\gamma$ of $(SA)_x(amine)_y$ (x=0-2, y=0-2) at 306 K, 1atm; cluster formation pathways for the SA−DMA at 286 K and 266 K; the ranges of temperature, [SA], [amine] and the data points data points for the input parameters of ACDC simulation; the detail observation data for time period, [SA], [DMA] in Beijing, Shanghai, Wangdu; the thermochemical information and cartesian coordinates of all clusters for $(SA)_x(amine)_y$ (x=0-2, y=0-2) system.

## AUTHOR INFORMATION

Corresponding Authors

Shuai Jiang − Tianjin Key Laboratory of Urban Transport Emission Research & State Environmental Protection Key Laboratory of Urban Ambient Air Particulate Matter Pollution Prevention and Control, College of Environmental Science and Engineering, Nankai University, Tianjin, 300071, China; ID: orcid.org/0000-0001-8015-4453

Jianfei Peng − Tianjin Key Laboratory of Urban Transport Emission Research & State Environmental Protection Key Laboratory of Urban Ambient Air Particulate Matter Pollution Prevention and Control, College of Environmental Science and Engineering, Nankai University, Tianjin, 300071, China; ID: orcid.org/0000-0003-4753-087X

Authors

Yongjian Lian – Tianjin Key Laboratory of Urban Transport Emission Research & State Environmental Protection Key Laboratory of Urban Ambient Air Particulate Matter Pollution Prevention and Control, College of Environmental Science and Engineering, Nankai University, Tianjin, 300071, China; ID: orcid.org/0009-0003-0330-689X

Xurong Bai – Tianjin Key Laboratory of Urban Transport Emission Research & State Environmental Protection Key Laboratory of Urban Ambient Air Particulate Matter Pollution Prevention and Control, College of Environmental Science and Engineering, Nankai University, Tianjin, 300071, China; ID: orcid.org/0009-0009-1168-1693

Ruoying Yuan - Tianjin Key Laboratory of Urban Transport Emission Research & State Environmental Protection Key Laboratory of Urban Ambient Air Particulate Matter Pollution Prevention and Control, College of Environmental Science and Engineering, Nankai University, Tianjin, 300071, China; ID: orcid.org/0009-0006-1909-4731

Wenli Xu - Tianjin Key Laboratory of Urban Transport Emission Research & State Environmental Protection Key Laboratory of Urban Ambient Air Particulate Matter Pollution Prevention and Control, College of Environmental Science and Engineering, Nankai University, Tianjin, 300071, China; ID: orcid.org/0009-0009-2083-6597

Yiyang Jiao- Tianjin Key Laboratory of Urban Transport Emission Research & State Environmental Protection Key Laboratory of Urban Ambient Air Particulate Matter Pollution Prevention and Control, College of Environmental Science and Engineering, Nankai University, Tianjin, 300071, China; ID: orcid.org/0009-0002-0035-5226

Hongjun Mao – Tianjin Key Laboratory of Urban Transport Emission Research & State Environmental Protection Key Laboratory of Urban Ambient Air Particulate Matter Pollution Prevention and Control, College of Environmental Science and Engineering, Nankai University, Tianjin, 300071, China; ID: orcid.org/0000-0001-8263-7571

Complete contact information is available

**Notes**

The authors declare no competing financial interest.

**Acknowledgments**

This research was funded by National Natural Science Foundation of China (42477111), the Fundamental Research Funds for the Central Universities of China (63253201, 63251191, 63243126), Natural Science Foundation of Tianjin Municipality (24JCYBJC01700) and Tianjin Key Research and Development Project (24YFXTHZ00070). The work was carried out at National Supercomputer Center in Tianjin, and the calculations were performed on Tianhe new generation supercomputer. This work is also supported by the Supercomputing Center of Nankai University (NKSC). We thank Qingzhu Zhang, Shengming Wang for sharing with us the cluster structures of SA-DEA and SA-MDEA.

**Data Availability Statement** All study data are included in the article and/or SI Appendix.